# Influence of surface conductivity on the apparent zeta potential of calcite

Shuai Li[1], Philippe Leroy[1*], Frank Heberling[2], Nicolas Devau[1], Damien Jougnot[3], Christophe Chiaberge[1]

[1] BRGM, French geological survey, Orléans, France.
[2] Institute for Nuclear Waste Disposal, Karlsruhe Institute of Technology, Karlsruhe, Germany.
[3] Sorbonne Universités, UPMC Univ Paris 06, CNRS, EPHE, UMR 7619 METIS, Paris, France.

[*]Corresponding author and mailing address:
Philippe Leroy
BRGM

3 Avenue Claude Guillemin
45060 Orléans Cedex 2, France
E-mail: p.leroy@brgm.fr
Tel: +33 (0)2 38 64 39 73
Fax: +33 (0)2 38 64 37 19

## Abstract

Zeta potential is a physicochemical parameter of particular importance in describing the surface electrical properties of charged porous media. However, the zeta potential of calcite is still poorly known because of the difficulty to interpret streaming potential experiments. The Helmholtz-Smoluchowski (HS) equation is widely used to estimate the apparent zeta potential from these experiments. However, this equation neglects the influence of surface conductivity on streaming potential. We present streaming potential and electrical conductivity measurements on a calcite powder in contact with an aqueous NaCl electrolyte. Our streaming potential model corrects the apparent zeta potential of calcite by accounting for the influence of surface conductivity and flow regime. We show that the HS equation seriously underestimates the zeta potential of calcite, particularly when the electrolyte is diluted (ionic strength ≤0.01 M) because of calcite surface conductivity. The basic Stern model successfully predicted the corrected zeta potential by assuming that the zeta potential is located at the outer Helmholtz plane, i.e. without considering a stagnant diffuse layer at the calcite-water interface. The surface conductivity of calcite crystals was inferred from electrical conductivity measurements and computed using our basic Stern model. Surface conductivity was also successfully predicted by our surface complexation model.

# 1. Introduction

The calcite-water interface has received ample attention during the past decades due to its high reactive properties and usefulness in many environmental and industrial applications [1, 2]. These applications include waste water purification [3], nuclear waste and $CO_2$ sequestration in geological formations [4-6], oil extraction [7], biomineralization [8], and cement and paper production [9, 10]. Furthermore, heavy metals and other contaminants can be adsorbed at calcite surface and be incorporated into the calcite crystal structure [11, 12]. Adsorption/desorption, dissolution, and precipitation phenomena at the calcite surface can be described by an electrostatic surface complexation model computing the behavior of the electrical double layer (EDL) at the calcite-water interface [1, 2].

Accurate acid-base potentiometric titration measurements of the surface charge of calcite cannot be easily performed because of the high reactivity of calcite in water [1, 2]. For that reason, electrokinetic experiments like electrophoresis or streaming potential are commonly performed to obtain reliable information on the structure of the calcite EDL [8, 11]. The streaming potential method is often used to characterize the electrochemical properties of calcite powders [9, 13]. During streaming potential experiments, the sample is subjected to a water pressure difference and the resulting water flow along the particles surface drags the excess of mobile charge of the pore water [14, 15]. A shear plane at the particles surface and a macroscopic electrical potential difference, the so-called streaming potential, appear during streaming potential experiments [16, 17]. The streaming potential method gives information on the electrical potential at the shear plane, i.e. on the zeta potential if conduction and streaming currents are correctly described [17, 18]. The zeta potential can be used to constrain the parameters (sorption equilibrium constants, capacitance(s)) of the electrostatic surface complexation model [1, 2].

However, the conversion of streaming potential measurements into zeta potentials is not straightforward because two effects, one associated with the surface conductivity of the material, and the other associated with the flow regime, decrease the streaming potentials [18, 19]. The Helmholtz-Smoluchowski (HS) equation neglects these two effects and its use can lead to underestimate zeta potentials [16, 20]. Heberling et al. [9, 10] used the HS equation to interpret their streaming potential experiments on a calcite powder in terms of apparent zeta potentials. Nevertheless, the HS equation can only be used when surface conductivity can be neglected and in the case of viscous laminar flow [16, 20]. Streaming potential induces electromigration currents in the EDL at the surface of the particles, which are responsible for surface conductivity [17, 18]. Surface conductivity increases conduction current opposed to streaming current and hence decreases the magnitude of the streaming potential [17, 19]. Inertial laminar flow decreases the apparent permeability, water flow in the pores and the resulting streaming potential [16, 21]. Given these observations, one may question whether the HS equation is appropriated to estimate the zeta potential of calcite powders from streaming potential experiments.

In the double layer theory, the zeta potential is considered to be located very close to the beginning of the diffuse layer [18, 19]. The viscosity of the diffuse layer is assumed to be equal to the viscosity of the bulk water and the liquid viscosity between the solid surface and the beginning of the diffuse layer is assumed to be significantly higher than the viscosity of the diffuse layer [19, 22]. Water flow along the particle surface is considered in the diffuse layer and bulk water and no water flow is considered between the solid surface and the beginning of the diffuse layer [18, 23]. This is the reason why it is assumed that the shear plane is located at the beginning of the diffuse layer in the double layer theory. Heberling et al. [9, 10] considered the presence of a stagnant diffuse layer at the calcite-water interface, i.e. a shear plane several nanometers away from the beginning of the diffuse layer, to reconcile high electrical potentials at the beginning of the diffuse layer computed by their surface complexation model to low measured apparent zeta potentials (the magnitude of the computed electrical potential decreases with the distance from the surface). Furthermore, on the contrary to silica where protruding polysilicic acid groups may increase the distance between the beginning of the diffuse layer and the shear plane [23, 24], there is no physical reason explaining the presence of a stagnant diffuse layer at the calcite surface. Heberling et al. [9, 10] also assumed that the thickness of the stagnant diffuse layer decreases with increasing salinity. This assumption is a typical signature of surface conductivity effects because the influence of surface conductivity on electrokinetic experiments decreases when salinity increases [25-29]. For instance, Heberling et al. [9, 10] assumed that the shear plane can be as far as 100–150 Å and 30–40 Å from the beginning of the diffuse layer at salinities of $10^{-3}$ M and $10^{-2}$ M NaCl, respectively. The location of the shear plane predicted by the surface complexation model of Heberling et al. [9, 10] is not in agreement with the double layer theory. Their use of the HS equation to interpret streaming potential experiments may explain why these authors considered low apparent zeta potentials and a large stagnant diffuse layer at low ionic strengths.

Revil and co-workers [14-16, 20] developed streaming potential models accounting for surface conductivity and Reynolds number effects. The Reynolds number is the ratio between inertial and viscous forces in the Navier–Stokes equation. Crespy et al. [20] successfully interpreted their streaming potential and electrical conductivity experiments on glass beads in contact with a NaCl solution in terms of low apparent and high corrected zeta potentials. Crespy et al. [20] showed considerably high surface conductivity and Reynolds number effects for glass beads pack immersed in a dilute electrolyte (salinity <0.01 M) and large glass beads (size >1000 μm), respectively. Nevertheless, Crespy et al. [20] did not use an electrostatic surface complexation model to interpret their streaming potential and conductivity measurements, thus their interpretation of streaming potentials in terms of surface complexation reactions is limited.

To the best of our knowledge, no study has used streaming potential, electrical conductivity measurements and an electrostatic surface complexation model to obtain the zeta potential and describe the behavior of the electrical double layer of calcite. After a brief theoretical description of the streaming potential, conductivity and surface complexation models, the zeta potentials of a calcite powder inferred from streaming potential and conductivity measurements are successfully reproduced by our basic Stern model (BSM). No assumption of a stagnant diffuse layer at the calcite surface is considered. Special care is given to the description of the surface processes responsible for the surface conductivity of calcite crystals.

## 2. Theoretical background

### 2.1. Streaming potential model

During streaming potential experiments, the sample is sandwiched between two water compartments, and the imposed water pressure difference induces a water flow and a shear plane at the particles surface [23, 30] (Fig. 1). The zeta potential is the electrical potential located at the shear plane [18, 19]. Water flow also drags the excess counter-ions in the diffuse layer along the pores surface and creates a macroscopic current density, the streaming current and a macroscopic electrical potential difference, the streaming potential [14, 17]. The electrical field induced by the streaming potential is responsible for conduction currents in the bulk pore water and in the EDL. The surface conductivity of the particles increases the conduction current density, which is opposed to the streaming current density [14, 18] and decreases the magnitude of the measured streaming potential (Fig. 1).

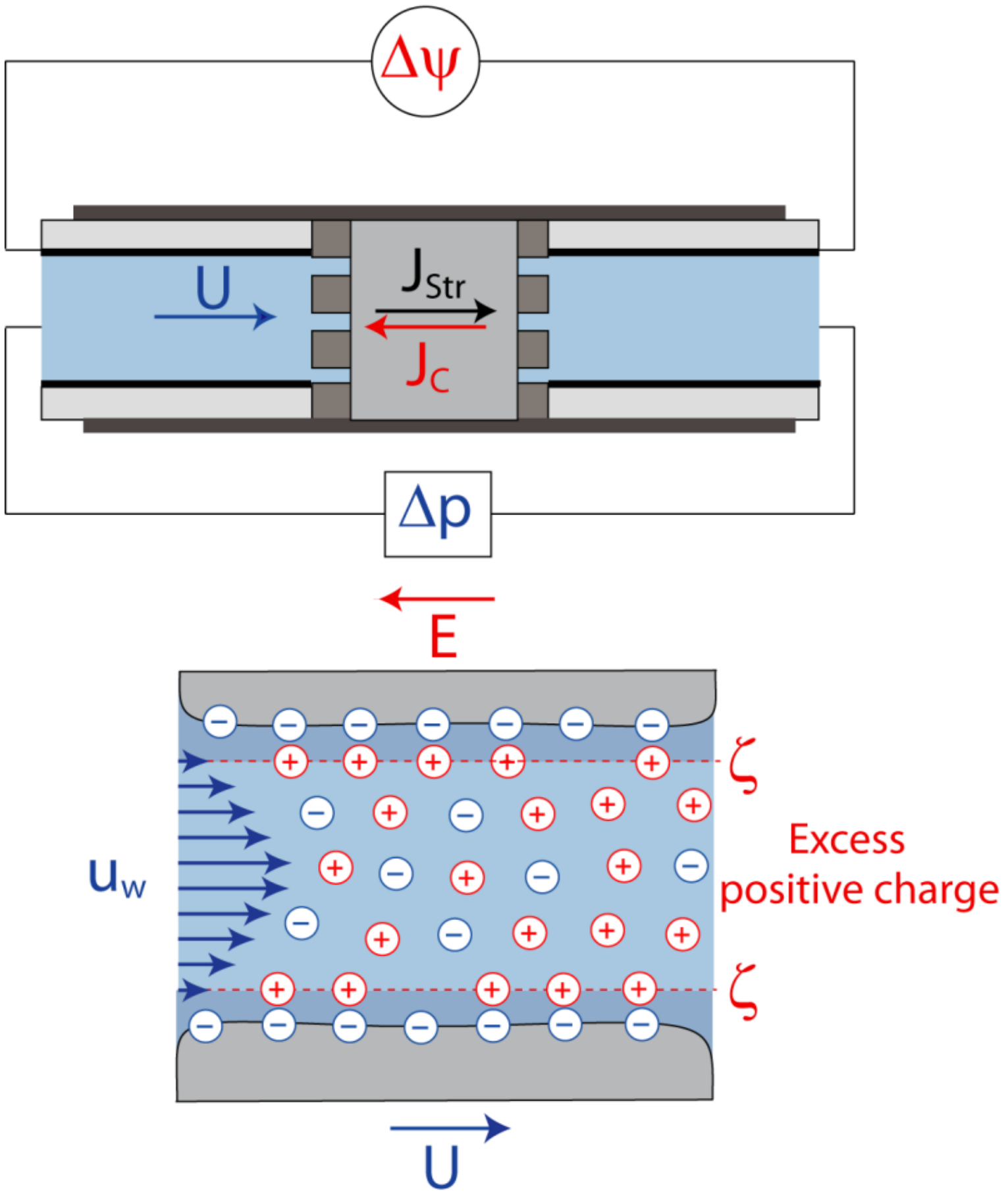

**Fig. 1.** Sketch of the streaming potential and of the electrokinetic coupling in a porous calcite sample. During streaming potential experiments, the local water velocity $\mathbf{u}_w$ due to the imposed water pressure difference $\Delta p$ drags the ions in the bulk water and diffuse layer along the pores surface. The displacement of the excess of charge in the diffuse layer is responsible for a shear plane at the particles surface and for a macroscopic electrical potential difference, the streaming potential $\Delta\psi$. The zeta potential ($\zeta$) is located at the shear plane. Ions displacement in the pores also induces two macroscopic current densities in opposite directions, the streaming current density $\mathbf{J}_{Str}$ (in A $m^{-2}$) and the conduction current density $\mathbf{J}_C$. The streaming current density results from the displacement of the mobile charge in the diffuse layer due to the water pressure difference. The conduction current density results from the displacement of the mobile charge in the electrical double layer and bulk water due to the streaming potential. The surface conductivity of the particles increases the conduction current density, which in turn decreases the magnitude of the measured streaming potential.

Revil et al. [14, 15], inspired by the work of Pride [31], used the volume averaging method to upscale the local Stokes and Poisson-Boltzmann equations at the scale of a representative elementary volume (REV). They gave the following equation describing the total current density due to conduction and streaming current densities in steady-state conditions and in the case of a macroporous material (thin EDL assumption) containing a viscous laminar flow:

$$\mathbf{J} = \mathbf{J}_{\mathbf{C}} + \mathbf{J}_{\mathbf{Str}} = -\sigma \nabla \psi + \frac{\varepsilon_w \zeta}{\eta_w F} \nabla p , \tag{1}$$

where $\sigma$ is the electrical conductivity of the porous medium (in S m$^{-1}$), $\psi$ is the macroscopic electrical potential (in V), $\varepsilon_w$ is the dielectric permittivity of water ($\varepsilon_w = \varepsilon_r \varepsilon_0$ where $\varepsilon_r$ is the relative dielectric permittivity of water, $\varepsilon_r \cong 78.3$ for bulk water at a pressure of 1 bar and a temperature $T$ of 298 K, and $\varepsilon_0$ is the dielectric permittivity of vacuum, $\varepsilon_0 \cong 8.854 \times 10^{-12}$ F m$^{-1}$ [32]), $\eta_w$ is the dynamic viscosity of bulk water (in Pa s; $\eta_w \cong 0.8903 \times 10^{-3}$ Pa s for pure water at $T$ = 298 K [32]), $F$ is the electrical formation factor, $\zeta$ is the zeta potential (in V), and $p$ is the water pressure (in Pa).

The conductivity of the porous medium can be modeled by considering that bulk and surface conductivity of the grains $\sigma_S$ act in parallel [33, 34]:

$$\sigma = \frac{\sigma_w}{F} + \sigma_S , \tag{2}$$

where $\sigma_w$ is the conductivity of bulk pore water.

According to Archie's first law [35], the electrical formation factor $F$ depends on the connected porosity $\phi$ and cementation exponent $m$:

$$F = \phi^{-m}. \tag{3}$$

When the streaming potential is measured at both sides of the sample in steady-state conditions, the total current density $\mathbf{J}$ is zero, and the streaming potential coupling coefficient $C$ can be measured [20, 36]. The streaming potential coupling coefficient (in V Pa$^{-1}$) is defined by the ratio of the streaming potential to the imposed water pressure difference in steady-state conditions:

$$C = \left.\frac{\Delta\psi}{\Delta p}\right|_{\mathbf{J}=\vec{0}}. \tag{4}$$

According to Revil et al. [14, 15], the streaming potential coupling coefficient in the case of viscous laminar flow can be calculated as a function of the zeta potential, electrical formation factor, and sample electrical conductivity by combining Eqs. (1) and (4):

$$C_0 = \frac{\varepsilon_w}{\eta_w F \sigma}\zeta. \tag{5}$$

The Helmholtz-Smoluchowski equation is a limiting case of the equation developed by Revil et al. [14, 15] for the streaming potential coupling coefficient when surface conductivity effects can be neglected. By combining Eqs. (2) and (5) and neglecting surface conductivity, we recover the HS equation for the streaming potential coupling coefficient:

$$C_{\mathrm{HS}} = \lim_{\sigma_S \to 0} C_0 = \frac{\varepsilon_w}{\eta_w \sigma_w}\zeta_a, \tag{6}$$

where $\zeta_a$ is the apparent zeta potential inferred from the measured streaming potential coupling coefficient and HS equation ( $\zeta_a = \zeta$ in absence of surface conductivity effects).

Equation (5) developed by Revil et al. [14, 15] is very close to the HS equation (Eq. (6)), except that the electrical conductivity of the bulk pore water $\sigma_w$ in the HS equation is replaced by the product of the electrical formation factor and electrical conductivity of the sample, $F\sigma$, in Eq. (5). According to Eqs. (5) and (6), the ratio between apparent zeta potential and zeta potential corrected for surface conductivity effects can be calculated as a function of the ratio of the bulk water conductivity to the product of the formation factor with the sample conductivity:

$$\frac{\zeta_a}{\zeta} = \frac{\sigma_w}{F\sigma}. \qquad (7)$$

By combining Eqs. (2) and (7), the ratio of the apparent to the corrected zeta potential can be expressed as a function of the product of the formation factor with the macroscopic Dukhin number $\mathrm{DU} = \sigma_S / \sigma_w$ [16, 37]:

$$\frac{\zeta_a}{\zeta} = \frac{1}{1 + F\dfrac{\sigma_S}{\sigma_w}} = \frac{1}{1 + F\,\mathrm{DU}}. \qquad (8)$$

For high ionic strengths $I$ (typically ≥0.1 M), grains surface conductivity can be neglected compared to bulk water conductivity [15, 33], hence DU → 0, and, according to Eq. (8), the apparent zeta potential can be equal to the zeta potential corrected for surface conductivity effects. For lower ionic strengths $I$, $F$DU may not be neglected, and the apparent zeta potential may be smaller (in magnitude) than the zeta potential corrected for surface conductivity (Eq. (8)). Note that Eq. (8) suggests that the surface conductivity effects on apparent zeta potential become more important when the formation factor increases, i.e. these effects are stronger for porous media having a low porosity (i.e. a high formation factor $F$, Eq. (3)) than for porous media containing the same materials (similar grains and water) but having a higher porosity (i.e. a lower formation factor).

The flow regime also influences the measured streaming potential and streaming potential coupling coefficient [38, 39]. The HS equation assumes viscous laminar flow, whereas inertial laminar flow may also occur during streaming potential experiments [20]. Revil and co-workers [16, 20] considered the effects of the Reynolds number in the case of inertial laminar flow on the streaming potential coupling coefficient (see Appendix A):

$$C = \frac{C_0}{1+\mathrm{Re}} = \frac{\varepsilon_w}{\eta_w F \sigma (1+\mathrm{Re})} \zeta , \tag{9}$$

$$\mathrm{Re} = \frac{1}{2}\left(\sqrt{1+c} - 1\right), \tag{10}$$

$$c = \frac{2\rho_w}{\alpha m \eta_w^2} \frac{d^3}{F(F-1)^3} \left(\frac{\Delta p}{l}\right), \tag{11}$$

where Re is the Reynolds number, $\rho_w$ is the water volumetric density (in kg m$^{-3}$, $\rho_w \cong 997$ kg m$^{-3}$ at a pressure of 1 bar, temperature $T$ of 298 K and in the case of dilute aqueous solutions of ionic strengths <1 mol L$^{-1}$ [32]), $d$ is the mean grain diameter, $\alpha$ is an empirical coefficient depending on the square of the cementation exponent ($\alpha = 32m^2$ [40]), and $l$ is the sample length (in m) during streaming potential experiments. Eqs. (9)–(11) apply only for viscous (Re <0.1) and inertial laminar flows (0.1 <Re <100) [16, 20] (for higher Reynolds numbers, water flow can be considered as turbulent). When the Reynolds number increases, the apparent permeability and water flow decreases [16, 21]. Therefore, in the viscous and inertial laminar flow regimes, when the Reynolds number increases, the streaming potential and streaming potential coupling coefficient decrease [16, 20].

The ratio between apparent zeta potential and zeta potential corrected for surface conductivity and Reynolds number effects can be calculated as a function of the bulk water conductivity, formation factor, sample conductivity, and Reynolds number by combining Eqs. (6) and (9):

$$\frac{\zeta_a}{\zeta} = \frac{\sigma_w}{F\sigma(1+\mathrm{Re})}. \qquad (12)$$

The surface conductivity and the Reynolds number decrease the ratio of the apparent to the corrected zeta potential according to Eqs. (2) and (12). These two effects can be easily corrected if streaming potential and electrical conductivity measurements are available. These corrections of the apparent zeta potential will be performed in section 3.1. The formation factor $F$ entering into Eqs. (9) and (12) will be estimated by dividing the water conductivity to the sample conductivity for salinities high enough to neglect surface conductivity effects on sample conductivity (Eq. (2)). The Reynolds number Re entering into Eqs. (9) and (12) will be estimated as a function of the formation factor, applied water pressure difference, sample length and mean grain diameter using Eqs. (10) and (11). The corrected zeta potential will be estimated in section 3.2 according to Eq. (9) and the measured streaming potential coupling coefficient, sample conductivity, and the estimated formation factor and Reynolds number. In section 3.2, the corrected zeta potential will be compared to the zeta potential predicted by our electrostatic surface complexation model by assuming that the shear plane is located at the beginning of the diffuse layer. In addition, an electrical conductivity model will be used in section 3.3 to estimate the surface conductivity of the calcite crystals from sample and water conductivity measurements.

The estimated surface conductivity will be compared to the surface conductivity computed from the surface complexation model in section 3.3. The model used to describe the electrical conductivity of the calcite sample is presented in section 2.2.

## 2.2. Electrical conductivity model

In this section, the electrical conductivity of the calcite sample $\sigma$ is calculated using the differential effective medium (DEM) theory [41-44]. This conductivity model considers that the pores are saturated with water and interconnected. The conductivity of the calcite sample is calculated according to the formation factor, the cementation exponent, and the conductivities of the bulk water and the calcite crystals $\sigma_s$ using the following equation:

$$\sigma = \frac{\sigma_w}{F}\left(\frac{1-\sigma_s/\sigma_w}{1-\sigma_s/\sigma}\right)^m, \qquad (13)$$

where the cementation exponent *m* of calcite crystals is constrained between 1.3 and 1.5 [45].

In section 3.1, the formation factor *F* entering into Eq. (13) will be estimated according to the ratio of the bulk water conductivity to the sample conductivity at salinities high enough to neglect surface conductivity (typically for salinities ≥0.1 M in the case of minerals other than clays [46]). According to Eq. (13), if the formation factor is determined, the surface conductivity of the calcite crystals can be directly estimated from the water and sample conductivity measurements. We make a distinction here between calcite grains and smaller calcite crystals located on the surface of calcite grains. Calcite grains have larger volumes than calcite crystals and are assumed to control the water flow in the sample whereas calcite crystals have a larger surface area-to-volume ratio than calcite grains and are assumed to control the surface conductivity of the porous medium. The parameter $\sigma_s$ in Eq. (13) is the surface conductivity of the calcite crystals and the parameter $\sigma_S$ defined previously in section 2.1 (Eq. (2)) is the macroscopic surface conductivity of the porous medium that is not directly linked to the surface conductivity of the calcite crystals.

The surface conductivity of the calcite crystals can also be computed using an electrostatic surface complexation model according to the following equation [47]:

$$\sigma_s = \frac{2}{a}\Sigma_s, \tag{14}$$

where $a$ is the mean radius (in m) and $\Sigma_s$ is the specific surface conductivity (in S) of the crystals, which can be computed by an electrostatic surface complexation model [25, 29, 36]. Eq. (14) considers that the calcite crystals are spherical. In reality, calcite crystals are mostly rhombohedral [1, 9], but the spherical particles assumption is a good first-order approximation.

We also make a distinction here between the macroscopic Dukhin number DU defined in section 2.1 as the ratio of the sample's macroscopic surface conductivity to the water conductivity whereas the microscopic Dukhin number is defined as the ratio of the specific surface conductivity of the calcite crystals to the product of their radius with the water conductivity [19]. In the following, the Dukhin number is referred to the microscopic Dukhin number of the calcite crystals. The Dukhin number will be estimated in section 3.3 to show the effects of the crystals' surface conductivity on streaming and apparent zeta potential. For spherical particles, according to Eq. (14), the Dukhin number can be calculated as a function of the ratio of the crystals' surface conductivity to bulk water conductivity [29]:

$$\mathrm{Du} = \frac{\Sigma_s}{a\sigma_w} = \frac{\sigma_s}{2\sigma_w}. \tag{15}$$

The specific surface conductivity (or conductance) $\Sigma_s$ represents the excess electrical conductivity integrated in the vicinity of a solid surface in reference to the electrical conductivity of bulk water [23, 29, 48]. The specific surface conductivity is due to the electromigration (superscript "*e*") of charged species and the associated electro-osmotic processes (superscript "*os*") at the solid-water interface [18, 19, 36]:

$$\Sigma_s = \Sigma_s^e + \Sigma_s^{os} = \int_0^{\chi_D} \left[\sigma(x) - \sigma_w + \Omega_d(x)\beta_{os}^d(x)\right] dx, \tag{16}$$

where $x$ is the local distance from the surface (in m), $\chi_D$ is the total thickness of the EDL (usually $\chi_D \cong 2\kappa^{-1}$ where $\kappa^{-1}$ is the Debye length in m [18]), $\Omega_d$ is the volume charge density (in C $m^{-3}$), and $\beta_{os}^d$ is the electro-osmotic mobility in the diffuse layer (in $m^2$ $s^{-1}$ $V^{-1}$). The viscous drag of the hydrated ions by the streaming potential is responsible for water flow in the diffuse layer. This water flow is called electro-osmosis and is responsible for an additional conductivity in the EDL. The first term of Eq. (16) represents the excess Ohmic conductivity in the vicinity of the surface and the second term of Eq. (16) represents the electro-osmotic conductivity in the diffuse layer.

The specific surface conductivity can be calculated as a function of the excess of charge and ionic mobilities at the solid surface [36, 49]. When calcite is in contact with an aqueous NaCl electrolyte at a given partial pressure of $CO_2$ ($pCO_2$), protons are adsorbed at the mineral surface, and different types of counter-ions, $Na^+$, $Ca^{2+}$, $Cl^-$, $HCO_3^-$, and $CO_3^{2-}$ are adsorbed in the EDL to compensate the surface charge [10]. The excess of mobile charge in the EDL can be directly computed by an electrostatic surface complexation model [25, 50, 51]. In our specific surface conductivity model, protons adsorbed at the calcium surface sites, counter-ions in the Stern layer, and counter and co-ions in the diffuse layer are assumed to contribute to the total specific surface conductivity of calcite crystals [23, 29, 48] (Fig. 2). The specific surface conductivity of the calcite crystals is calculated using the following equations:

$$\Sigma_s = \Sigma_s^0 + \Sigma_s^{St} + \Sigma_s^d, \tag{17}$$

$$\Sigma_s^0 = e\beta^0_{\mathrm{H_3O^+}}\left(\Gamma^0_{\mathrm{CaO^{-1.5}-H_3O^+}} + 2\,\Gamma^0_{\mathrm{CaO^{-1.5}-2\,H_3O^+}}\right), \tag{18}$$

$$\Sigma_s^{St} = e\sum_{i=1}^{M} z_i \beta_i^{St}\,\Gamma_i^{St}, \tag{19}$$

$$\Sigma_s^d = \int_{x=x_d}^{x=x_d+\chi_D}\left[\sigma_d(x) - \sigma_w + \Omega_d(x)\beta_{os}^d(x)\right]dx, \tag{20}$$

where $\Sigma_s^0$, $\Sigma_s^{St}$ and $\Sigma_s^d$ are the respective contributions of the mineral surface (superscript "0"), Stern (superscript "*St*") and diffuse layer (superscript "*d*") to the total specific surface conductivity. The parameter $e$ is the elementary charge (of value ≅ 1.602×10$^{-19}$ C), $\beta^0_{\mathrm{H_3O^+}}$ is the hydronium mobility along the mineral surface, and $\Gamma^0_{\mathrm{CaO^{-1.5}-H_3O^+}}$ and $\Gamma^0_{\mathrm{CaO^{-1.5}-2\,H_3O^+}}$ are the surface site densities of adsorbed hydronium ions at the mineral surface (in sites m$^{-2}$) (including the surface sites where ions are adsorbed in the Stern layer). The parameter $M$ is the number of adsorbed species in the Stern layer, $z_i$ is the ion valence, $\beta_i^{St}$ is their mobility, and $\Gamma_i^{St}$ is their surface site density in the Stern layer. The parameter $x_d$ is the distance of the beginning of the diffuse layer from the mineral surface, and $\sigma_d$ is the electrical conductivity of the diffuse layer.

Hydrophobic media possessing low dielectric permittivity like Teflon [52], air [27], or amorphous silica [29] are characterized by large hydration layers at their surface, which may allow the migration of hydrated protons (hydronium ions) along the particle surface via the hydrogen bonding network. We assume here that the calcite-water interface, also characterized by large hydration layers, allows the hydrated protons to move along the particle surface. It is interesting to note that Holmes et al. [53] considered the specific surface conductivity of ionizable surface hydroxyl groups at the surface of thorium oxide, and that Revil and co-workers [23, 48], Zimmermann et al. [52], and Leroy and co-workers [27, 29] considered the contribution of protons to the specific surface conductivity of silica, Teflon, and air and silica, respectively.

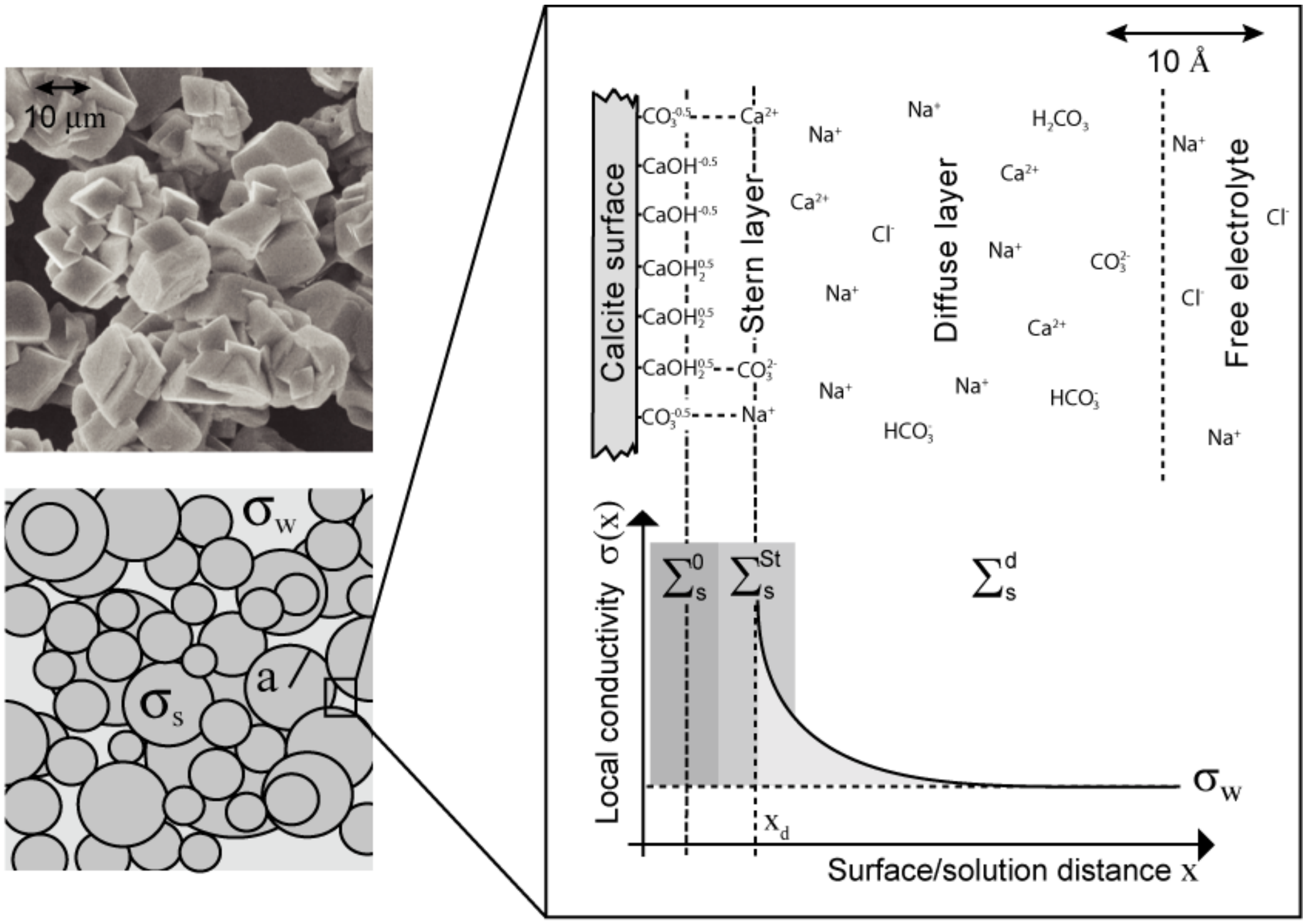

**Fig. 2.** The electrochemical properties of calcite in contact with an aqueous NaCl solution and atmospheric $CO_2$. Each calcite crystal has a local excess of electrical conductivity at the mineral surface due to the mobile protons, and in the Stern and diffuse layer due to the electromigration

of counter-ions and co-ions under the streaming potential. The specific surface conductivity of the diffuse layer is estimated by integrating $\sigma(x)-\sigma_w$ over its entire thickness.

Electromigration currents in bulk water are responsible for its electrical conductivity. The bulk water conductivity $\sigma_w$ can be calculated according to the following equation [54]:

$$\sigma_w = e1000\,\mathrm{N_A}\sum_{i=1}^{N} z_i \beta_i^w C_i^w\,, \tag{21}$$

where $\mathrm{N_A}$ is the Avogadro number (of value ≅ $6.022\times10^{23}$ mol$^{-1}$), $N$ is the number of types of ions in the bulk pore water, $\beta_i^w$ is the mobility (in m$^2$ s$^{-1}$ V$^{-1}$) and $C_i^w$ is the concentration (in mol dm$^{-3}$, M) of ion $i$ in the bulk water (superscript "$w$").

The specific surface conductivity of the diffuse layer is computed considering the effects of electromigration and electro-osmosis, by using Eqs. (20), (21) and the resulting equation, which is [51]:

$$\Sigma_s^d = e1000\,\mathrm{N_A}\sum_{i=1}^{N} z_i C_i^w \int_{x=x_d}^{x=x_d+\chi_D}\left\{B_i^d(x)\exp\left[-\frac{q_i\varphi(x)}{k_bT}\right]-\beta_i^w\right\}dx\,, \tag{22}$$

$$B_i^d(x) = \beta_i^d(x) + \pm\beta_{os}^d(x) \cong \beta_i^w + \pm\frac{\varepsilon_w}{\eta_w}\left[\varphi(x)-\varphi_d\right], \tag{23}$$

$$\kappa^{-1} = \sqrt{\frac{\varepsilon_w k_b T}{2e^2 1000\,\mathrm{N_A}\, I}}\,, \tag{24}$$

$$I = 0.5\sum_{i=1}^{N} z_i^2 C_i^w , \tag{25}$$

where $B_i^d$ is the ion effective mobility in the diffuse layer (in m$^2$ s$^{-1}$ V$^{-1}$) including an electro-osmotic contribution, $q_i = \pm e z_i$ is the ion charge ("+" stands for cations and "–" stands for anions), $k_b$ is the Boltzmann constant (of value ≅ 1.381×10$^{-23}$ J K$^{-1}$), $T$ is the temperature (in K), $\varphi$ is the electrical potential in the diffuse layer (in V), and $\beta_i^w$ is the ion mobility in bulk water (in m$^2$ s$^{-1}$ V$^{-1}$). In Eqs. (23) and (24), the ion mobility, water dielectric permittivity, and viscosity in the diffuse layer are assumed to be equal to their values in bulk water, and $\varphi_d$ is the electrical potential at the beginning of the diffuse layer. Eq. (22) is equivalent to the Bikerman's equation for the calculation of the specific surface conductivity (or conductance) of the diffuse layer generalized for any kind of ion in the diffuse layer [36, 48, 51]. This equation considers an electro-osmotic contribution to the specific surface conductivity of the diffuse layer. In Eq. (23), electro-osmosis increases the effective mobilities of the counter-ions and decreases the effective mobilities of the co-ions (less numerous than the counter-ions), hence electro-osmosis increases the specific surface conductivity of the diffuse layer. The contribution of electro-osmosis to the specific surface conductivity of the diffuse layer can be as high as 50% of the total specific surface conductivity of the diffuse layer [51, 55]. In Eqs. (22) and (23), the electrical potential distribution in the diffuse layer $\varphi(x)$ is computed by numerically solving the Poisson-Boltzmann equation (see Eqs. (16)–(19) in Leroy et al. [51] for further details concerning the computation procedure). Note that spherical coordinates are not considered for the calculation of the specific surface conductivity of the diffuse layer, because calcite crystals are considered larger than the thickness of the diffuse layer for the experimental conditions investigated here.

The surface conductivity model presented here also implies that the shear plane is located at the beginning of the diffuse layer because no water flow is assumed in the Stern layer (no electro-osmosis is assumed in the Stern layer). We assume that the mineral surface and the Stern layer behave like a gel where only ions can move along the particle surface under the influence of the streaming potential [22]. A big unknown of the surface conductivity model are the values of the surface mobilities of the protons and of the counter-ions in the Stern layer. The comparison of the surface conductivity of the calcite crystals estimated from conductivity measurements and predicted by the surface complexation model can give precious information on the surface mobilities of the protons at the mineral surface and counter-ions in the Stern layer, and on the contributions of the mineral surface, Stern, and diffuse layers to the surface conductivity of the calcite crystals. This comparison will be discussed in section 3.3.

The following section will briefly introduce the electrostatic surface complexation model used to compute the surface electrical properties of calcite. It was developed by Heberling et al. [9, 10].

## 2.3. Calcite surface complexation model

For most of the oxide minerals, surface adsorption is mainly controlled by one hydroxylated metal cation, $> \text{MeOH}$ surface site, resulting from the hydrolysis of adsorbed water molecules [11, 57]. Calcite surface functional groups of the dominating crystallographic plane (calcite (1 0 4) surface) behave differently than those of oxides, because they are assumed to be controlled by two different types of sites resulting from the hydrolysis of surface water molecules: a hydroxylated calcium cation, $>CaOH$, and a protonated carbonate anion, $>CO_3H$ surface site [8, 11]. This assumption of hydrolysis of surface water molecules at the calcite-water interface was recently questioned by a theoretical study showing that surface water molecules may not be dissociated and that $>CaOH_2$ and $>CO_3$ surface sites may thus control the calcite surface speciation [58]. Following the recent surface complexation model of Heberling et al. [10], we will also consider the presence of $>CaOH_2$ and $>CO_3$ surface sites at the calcite surface.

The calcite surface is defined by the position of the surface $Ca^{2+}$ ions [11] and the "0-plane", where adsorption of protons occurs, is considered to be located a few Ångströms (between 1.2 [1] and 2.3 Å [10]) from the calcite surface (Fig. 3). According to Heberling et al. [9], the Stern plane, the "$\beta$-plane" is located beyond the two hydration water layers at a distance of 4–6 Å from the calcite surface. Therefore, counter-ions in the Stern layer are assumed to be mostly adsorbed as outer-sphere surface complexes at a distance of a few Ångströms from the "0-plane" [8]. In our model, calcium and sodium ions are assumed to be adsorbed in the same plane, despite a recent study showing, using molecular dynamics (MD) simulations and atomic force microscopy (AFM) measurements, that sodium ions are located closer to the calcite surface than calcium ions due to their smaller hydration shell [59]. Nevertheless, only streaming potential experiments of a calcite powder in contact with an aqueous NaCl solution and a low $pCO_2$ ($pCO_2=10^{-3.44}$ atm) will be investigated in our study. Given the low $Ca^{2+}$ concentration in the pore water, our assumption about a common adsorption plane for the different counter-ions in the Stern layer will not influence the predicted electrochemical properties.

In this study, only the surface complexation reactions at the calcite (1 0 4) surface, which is the dominating crystallographic plane on most types of calcite, are considered. The basic Stern model developed recently by Heberling et al. [10] is used to describe the electrochemical properties of the mineral surface, Stern, and diffuse layer of the calcite-water interface. The basic Stern model considers that the Stern plane, the "$\beta$-plane" coincides with the "$d$-plane", i.e. $\varphi_\beta = \varphi_d$ (Fig. 3). According to crystallographic studies, the total surface site density of calcium and carbonate surface sites on the calcite (1 0 4) face was estimated to be equal to 4.95 sites per nm$^{-2}$ for each type of site [1, 9, 10]. Heberling et al. [10] assumed that the calcite surface functional groups are controlled by the $> CaOH_2^{0.5}$, $> CaOH^{-0.5}$ and $> CO_3^{-0.5}$ surface sites. Heberling et al. [10] also assumed that the surface charge of their calcite samples is negative as a whole in the investigated pH range [5.5, 10.5]. Therefore, in our surface complexation model, the calcite surface charge is compensated by hydrated cations in higher concentrations at the Stern layer than hydrated anions.

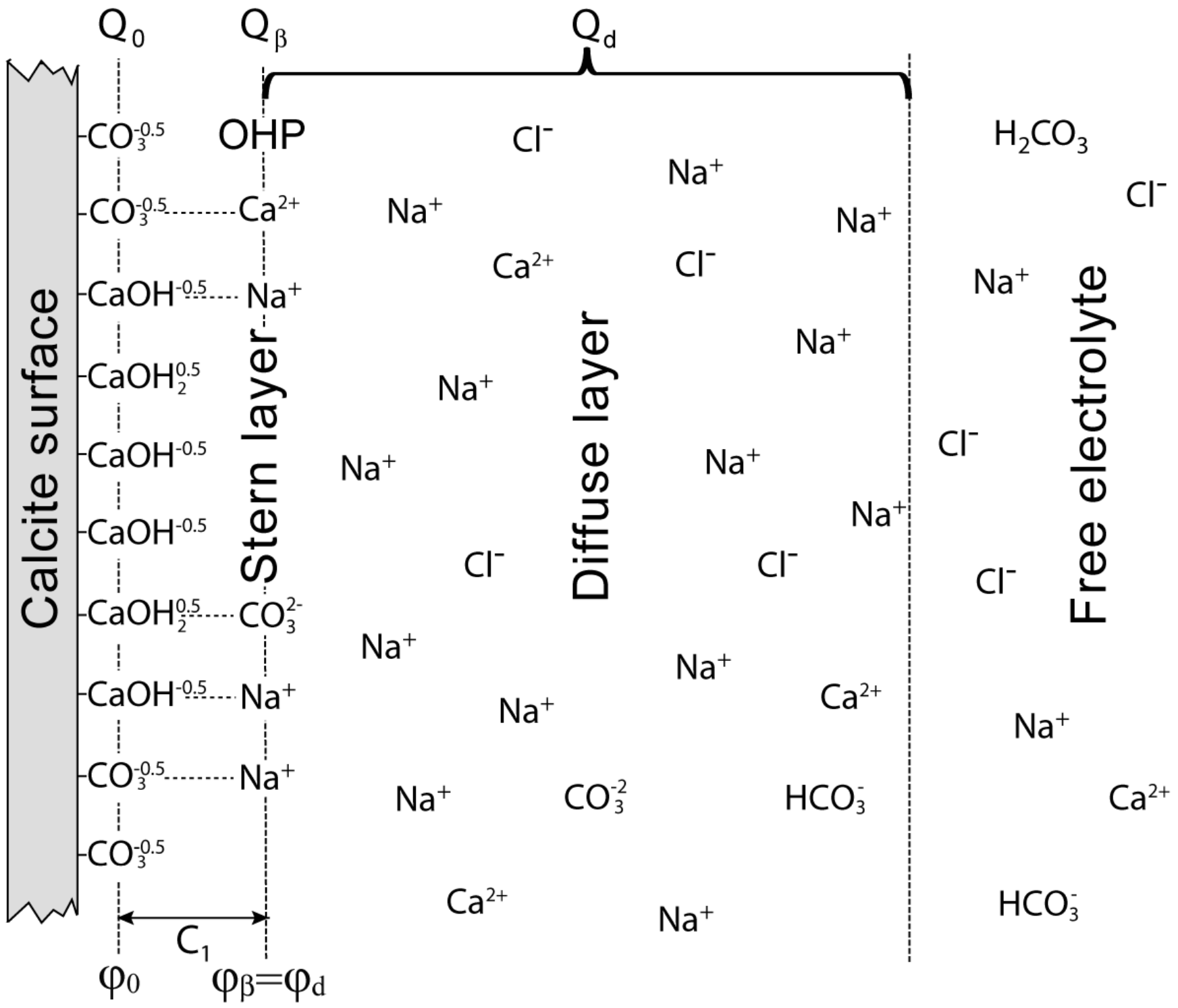

**Fig. 3.** The basic Stern model used by Heberling et al. [10] to describe the calcite-water interface (calcite (1 0 4) surface) when calcite is in contact with an aqueous NaCl solution and gaseous $CO_2$. Calcium and carbonate ions come from the reactions of calcite with water and water with the atmosphere. The parameters $\varphi$ and $Q$ are the electrical potential and surface charge density, respectively, at the "0-plane", "$\beta$-plane" (Stern plane) and "$d$-plane" (beginning of the diffuse

layer) ($Q_d$ is the surface charge density of the diffuse layer). According to the basic Stern model, the beginning of the diffuse layer coincides with the Stern plane, i.e. $\varphi_\beta = \varphi_d$. OHP is the Outer Helmholtz plane. The parameter $C_1$ is the Helmholtz capacitance to model the electrical potential behavior between the "0-plane" and "$\beta$-plane".

A geochemical code written in Matlab was developed to compute the surface complexation model of Heberling et al. [10] (see Appendix B). The program combines aqueous complexation, surface complexation equilibria, surface charge density and mass balance conditions [11, 60]. The influence of acid (HCl) and base (NaOH) on the pH of the aqueous solution and electrical properties of the calcite-water interface is considered [61]. The set of equations obtained is solved iteratively by the classical Newton-Raphson technique [62]. Table 1 summarizes, in matrix form, the set of equations used for the calculation of surface and solution speciation for fixed values of pH and $pCO_2$.

On the contrary to Heberling et al. [10], our Matlab code does not consider a stagnant diffuse layer at the calcite-water interface, i.e, the zeta potential ($\zeta$) is directly computed assuming that the shear plane is located at the Stern plane, i.e. $\varphi_\beta = \varphi_d = \zeta$. This assumption is in agreement with the EDL theory: no water flow is considered in the Stern layer and water flow is considered in the diffuse layer. In section 3.2, only the capacitance $C_1$ of the surface complexation model will be optimized to match the corrected zeta potentials inferred from the streaming and conductivity measurements on a calcite powder.

**Table 1.** Stoichiometric matrix of aqueous and surface reactions at the calcite-water interface. The parameters $\varphi$ and $K$ are the electrical potentials at the different planes and the equilibrium constants of the reactions, respectively (adapted from the database Phreeqc.dat of the geochemical software Phreeqc [56] and from Heberling et al. [10]).

| Product species | $H^+$ | $Cl^-$ | $Na^+$ | $Ca^{2+}$ | $HCO_3^-$ | $>CaOH^{-0.5}$ | $>CO_3^{-0.5}$ | $e^{\frac{-e\varphi_0}{k_bT}}$ | $e^{\frac{-e\varphi_\beta}{k_bT}}$ | $\log_{10}K$ |
|---|---|---|---|---|---|---|---|---|---|---|
| $CO_3^{-2}$ | -1 | 0 | 0 | 0 | 1 | 0 | 0 | 0 | 0 | -10.33 |
| $H_2CO_3$ | 1 | 0 | 0 | 0 | 1 | 0 | 0 | 0 | 0 | 6.35 |
| $CaHCO_3^+$ | 0 | 0 | 0 | 1 | 1 | 0 | 0 | 0 | 0 | 1.11 |
| $CaCO_3(aq)$ | -1 | 0 | 0 | 1 | 1 | 0 | 0 | 0 | 0 | -7.10 |
| $CaOH^+$ | -1 | 0 | 0 | 1 | 0 | 0 | 0 | 0 | 0 | -12.78 |
| $>CaOH_2^{+0.5}$ | 1 | 0 | 0 | 0 | 0 | 1 | 0 | 1 | 0 | 0.50 |
| $>CaOH_2^{+0.5}\cdots Cl^-$ | 1 | 1 | 0 | 0 | 0 | 1 | 0 | 1 | -1 | 0.45 |
| $>CaOH_2^{+0.5}\cdots HCO_3^-$ | 1 | 0 | 0 | 0 | 1 | 1 | 0 | 1 | -1 | 0.54 |
| $>CaOH_2^{+0.5}\cdots CO_3^{2-}$ | 0 | 0 | 0 | 0 | 1 | 1 | 0 | 1 | -2 | -6.57 |
| $>CaOH^{-0.5}\cdots Na^+$ | 0 | 0 | 1 | 0 | 0 | 1 | 0 | 0 | 1 | 0.56 |
| $>CaOH^{-0.5}\cdots Ca^{2+}$ | 0 | 0 | 0 | 1 | 0 | 1 | 0 | 0 | 2 | 1.68 |
| $>CO_3^{-0.5}\cdots Na^+$ | 0 | 0 | 1 | 0 | 0 | 0 | 1 | 0 | 1 | 0.56 |
| $>CO_3^{-0.5}\cdots Ca^{2+}$ | 0 | 0 | 0 | 1 | 0 | 0 | 1 | 0 | 2 | 1.68 |

# 3. Comparison with experimental data and discussion

## 3.1. Apparent and corrected zeta potentials

The streaming potential model presented in section 2.1 is used to interpret the streaming potential and electrical conductivity measurements on a calcite powder immersed in an aqueous NaCl solution (salinities = $10^{-3}$, $10^{-2}$, $5\times10^{-2}$ M, pH range = [5.5, 10.5]) and in contact with the atmosphere ($pCO_2=10^{-3.44}$ atm) (Fig. 4). The measurements were performed using an Anton Paar SurPASS electrokinetic analyzer [9]. Coarse crystallites >25 µm in diameter were ground from natural calcite single crystals var. Iceland spar from Chihuahua Mexico, that have been purchased from Ward's Natural Science.

The measured streaming potential coupling coefficient is negative due to the negative surface charge of calcite in the investigated pH range and the magnitude of the coupling coefficient decreases when salinity increases due to the compression of the electrical diffuse layer and the increase of the conductivity of the pore water when salinity increases (Fig. 4). The curves showing the evolution of the measured streaming potential coupling coefficient and calcite conductivity with pH are relatively flat, except for the lowest salinity (1 mM NaCl) because of the influence of the concentration of acid (HCl), base (NaOH), and dissolved $CO_2$ on the water chemical composition [61, 63]. At a partial pressure of $CO_2$ of $10^{-3.44}$ atm, the concentration of $CO_2$ dissolved in water is rather small and is equal to $1.487\times10^{-5}$ M. Therefore, the concentrations of carbonates, $CO_3^{2-}$, and bicarbonate ions, $HCO_3^-$, will not influence significantly the surface speciation of our calcite sample.

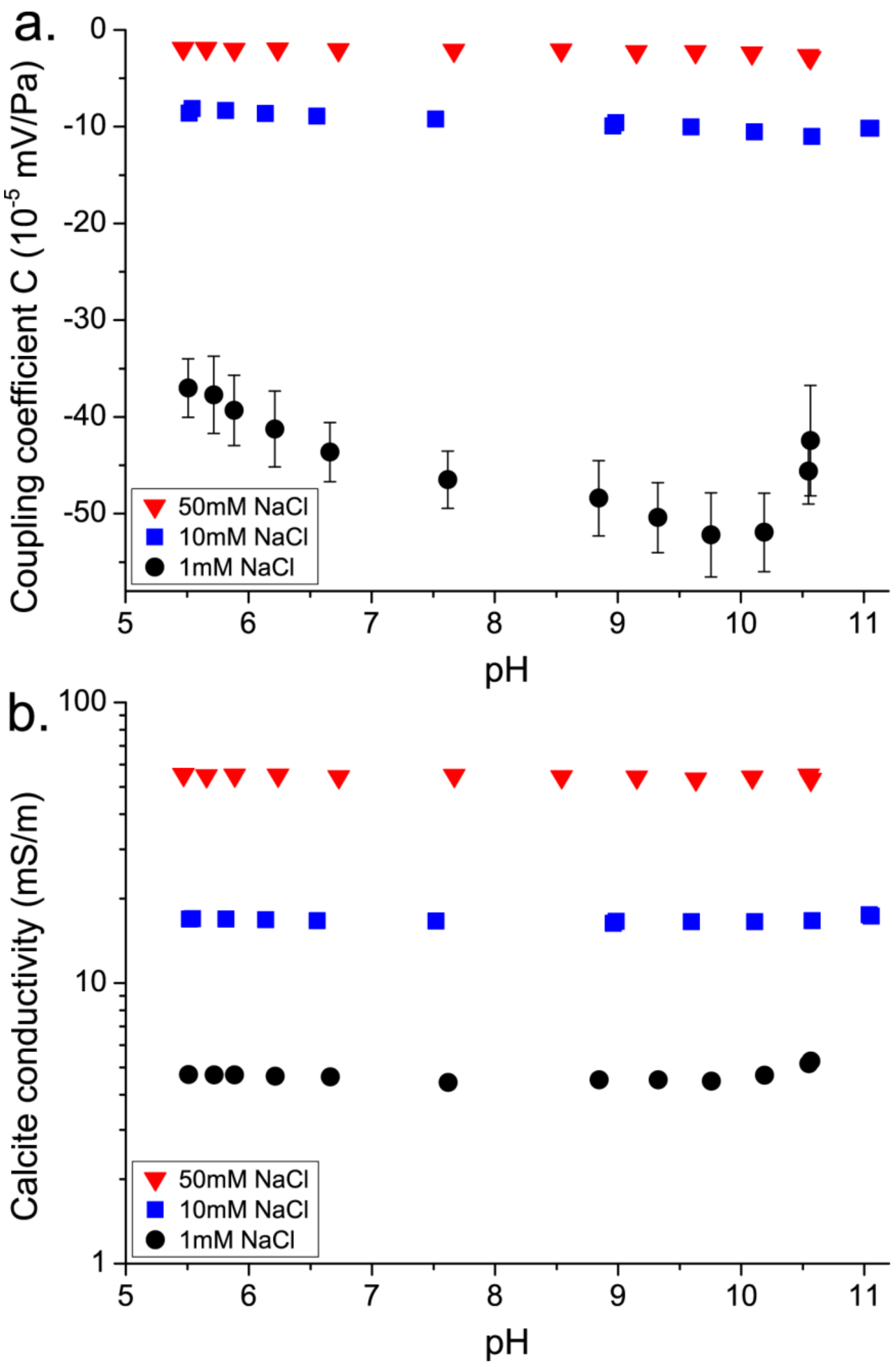

**Fig. 4.** Measured streaming potential coupling coefficient (a.) and sample conductivity (b.) versus pH for different water conductivities, equal to 28, 135 and 580 mS $m^{-1}$, corresponding to salinities equal to 0.001, 0.01 and 0.05 M NaCl, respectively. The error bars represent the streaming potential coupling coefficients measured during four pumping cycles (twice back and forth). For the two highest salinities, the error bars are smaller than the sizes of the symbols.

Sample conductivity measurements were also performed at different salinities and fixed pH (pH = 9.5) to estimate the electrical formation factor. The electrical formation factor $F$ is a parameter of paramount importance in our streaming potential model because it considers the effects of the porosity and particles shape on electrical measurements. This parameter was calculated by fitting the measured apparent formation factor $F_a = \sigma_w / \sigma$ using the DEM model (Eq. (13) with $m$ = 1.5), adjusted surface conductivities of the calcite crystals $\sigma_s$ and water conductivity measurements $\sigma_w$. By extrapolating the DEM model to very high salinities (water conductivities > 1 S m$^{-1}$), we found a value of 13.5 for the intrinsic formation factor, corresponding to a connected porosity of 0.18 using Archie's law ($F = \phi^{-m}$) (Fig. 5).

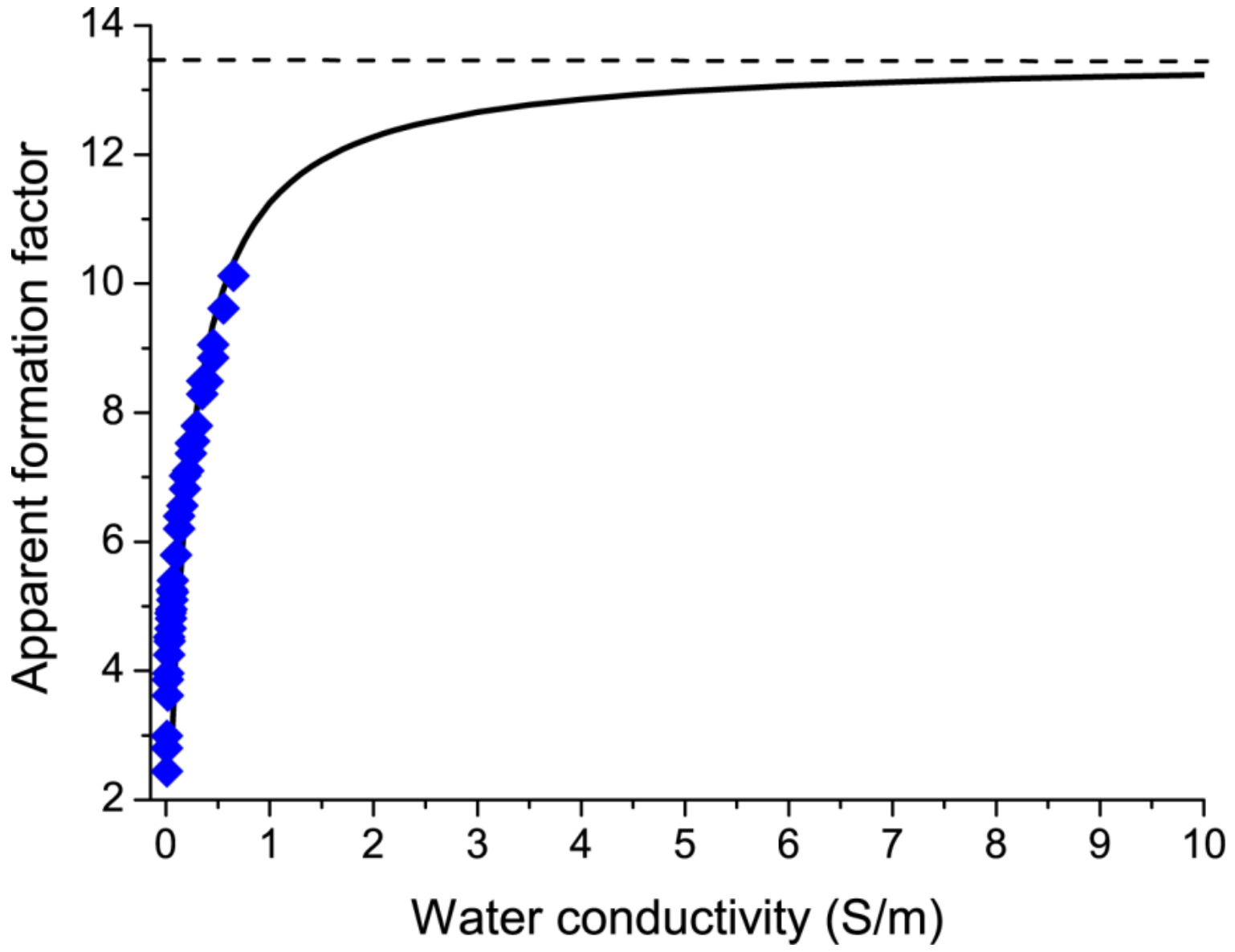

**Fig. 5.** Apparent formation factor of a calcite powder as a function of water conductivity (aqueous NaCl solution). Symbols are measurements and the line is the predictions of the DEM model. The intrinsic formation factor is estimated by extrapolating the apparent formation factor predicted by the DEM model for the highest water conductivities. The determined intrinsic formation factor *F* is 13.5.

The Reynolds number was then calculated using the intrinsic formation factor, Eqs. (10), (11), the measured averaged water pressure difference ($\Delta p$ = 275 mbar), the length of the sample (2 cm) and the measured averaged grain diameter ($d$ = 550 μm). We found Re = 0.1, hence the regime of the water flow is between inertial and viscous laminar, and the impact of the Reynolds number on streaming potential measurements is rather small. Eq. (12) is used to calculate the ratio of the apparent to the corrected zeta potential as a function of the formation factor, Reynolds number, and measured water and sample conductivities for different pH and salinities (Fig. 6).

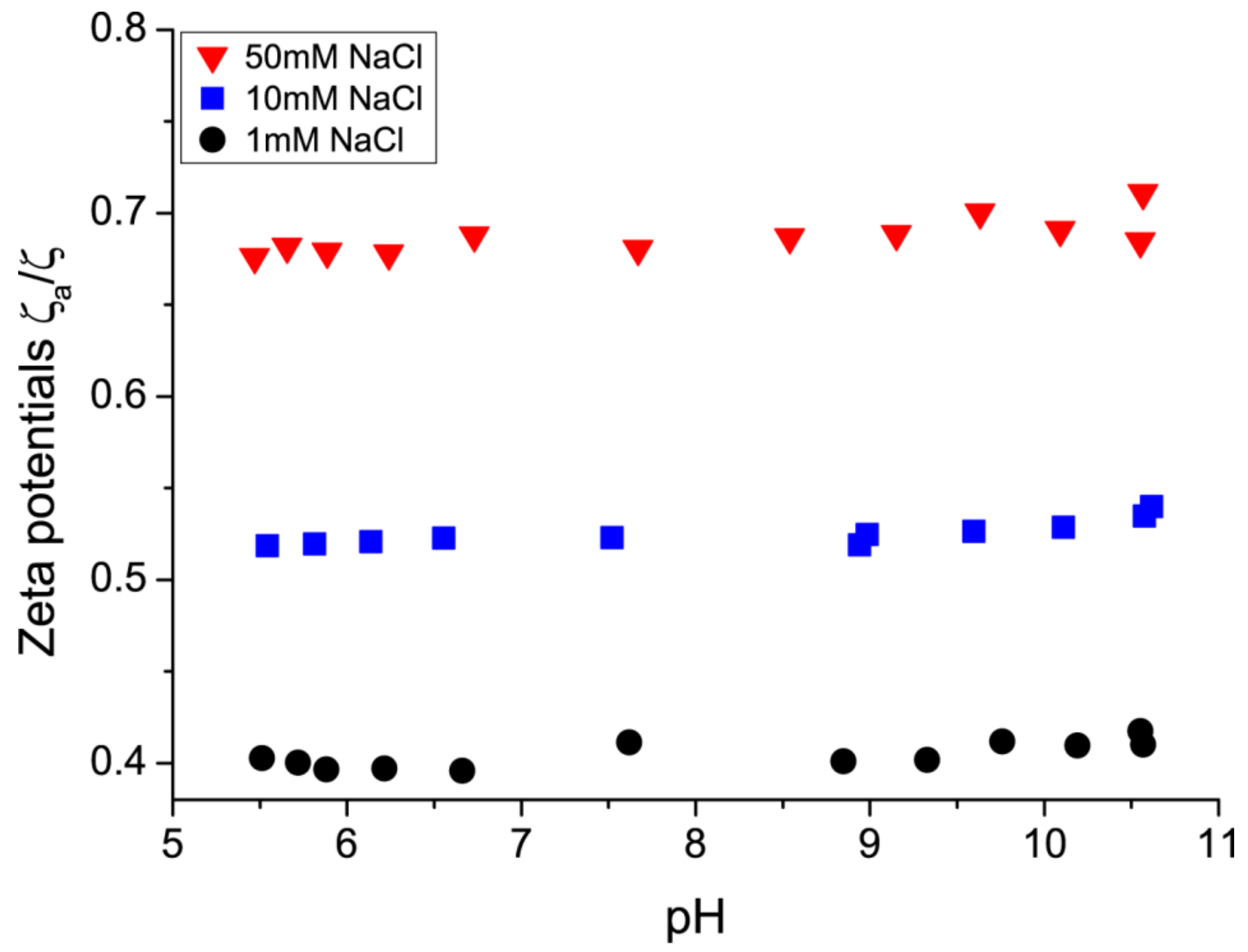

**Fig. 6**. Ratios of the apparent to corrected zeta potential versus pH at different salinities (0.001, 0.01 and 0.05 M NaCl).

The zeta potential, corrected for surface conductivity and Reynolds number effects, is considerably higher (in magnitude) than the apparent zeta potential inferred from the HS equation, the ratio of the apparent to corrected zeta potential is smaller than 0.7 whatever the pH and salinity (Fig. 6). The ratio of the apparent to corrected zeta potential decreases with the dilution of the aqueous electrolyte, from approximately 0.7 at a salinity of 0.05 M NaCl to approximately 0.4 at a salinity of 0.001 M NaCl. The corrected zeta potential is therefore approximately 2.5 times the apparent zeta potential at a salinity of $10^{-3}$ M NaCl. These results demonstrate that the surface conductivity of calcite crystals can't be neglected for our experimental conditions and that surface conductivity considerably increases conduction current, which decreases the magnitude of the streaming potential and apparent zeta potential. It is interesting to note that the zeta potentials ratios do not depend on pH, i.e. the effects of surface conductivity on streaming potential measurements are not sensitive to pH, in other words, surface conductivity of calcite crystals is not sensitive to pH. Surface conductivity will be examined in details in section 3.3 using our surface complexation model. In the next section, the corrected zeta potentials will be compared to the zeta potentials predicted by our basic Stern model.

### 3.2. Zeta potential predicted by the basic Stern model

Our surface complexation model was used to compute the electrical potential at the Stern plane, and this potential was compared to the corrected experimental zeta potential. The capacitance $C_1$ describing the electrical potential behavior between the "0-plane" and the "$\beta$-plane" was adjusted using the Simplex algorithm [64] to match the modeled to the observed zeta potentials. This algorithm decreases in a least square sense the cost function between observed and computed zeta potentials (read Caceci and Cacheris [64] and Leroy and Revil [50] for more information regarding the Simplex algorithm). Other surface complexation parameters (equilibrium sorption constants) were taken from Heberling et al. [10] (Table 1).

The predicted zeta potentials are in very good agreement with the experimental data without assuming a stagnant diffuse layer at the calcite-water interface (Fig. 7). The zeta potential data at different salinities can easily be reproduced by our basic Stern model. The apparent zeta potential data are also showed by comparison (Fig. 7). On the contrary to the corrected zeta potential data, the apparent zeta potential data at different salinities can't be easily distinguished because they are not corrected of surface conductivity effects and their magnitudes are considerably smaller than the magnitudes of corrected zeta potentials, in particular for low salinities (1 and 10 mM NaCl). Our treatment therefore considerably improves the interpretation of streaming potential measurements and is able to reconcile experimental zeta potentials to zeta potentials inferred from a surface complexation model without assuming a stagnant diffuse layer. In addition, our calculations show that the Helmholtz-Smoluchowski equation is not suitable to estimate the zeta potential of calcite powders from streaming potential measurements because it neglects the surface conductivity of the calcite crystals.

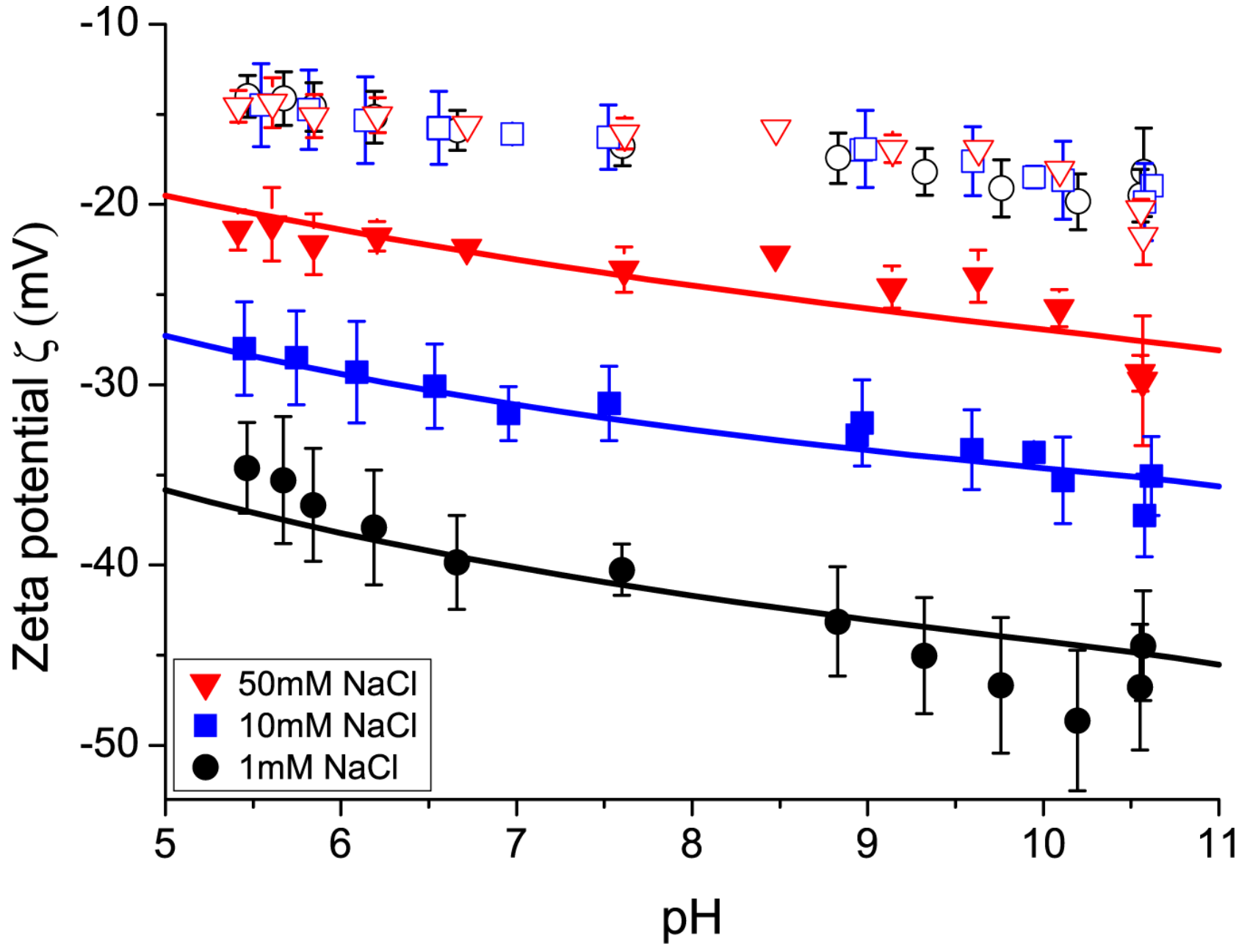

**Fig. 7.** Calcite zeta potential versus pH at different salinities (0.001, 0.01 and 0.05 M NaCl). Full symbols represent the corrected experimental zeta potentials and lines are the predictions of the basic Stern model. Empty symbols are the apparent zeta potentials derived from the Helmholtz-Smoluchowski equation. The error bars represent the zeta potential data from the streaming potential measurements during four pumping cycles.

We found a capacitance $C_1$ equal to 1.24 F m$^{-2}$, which is higher than the value of $C_1$ equal to 0.45 F m$^{-2}$ reported by Heberling et al. [10]. The capacitance $C_1$ can be described by the following equation:

$$C_1 = \frac{\varepsilon_1}{x_1}, \tag{26}$$

where $\varepsilon_1$ is the water dielectric permittivity and $x_1$ is the distance between the "0-plane" and the "$\beta$-plane".

This higher value of $C_1$ compared to the value reported by Heberling et al. [10] suggests that sodium ions may not be located beyond the two hydration layers, but closer to the surface, as observed by Ricci et al. [59] (smaller $x_1$) and/or that the dielectric permittivity $\varepsilon_1$ between the "0-plane" and the "$\beta$-plane" is larger than the value of $12\varepsilon_0$ reported by Heberling et al. [10] (Eq. (26)). Most importantly, the values for the Helmholtz capacitance, $C_1$, or in turn the Stern layer thickness, $x_1$, and permittivity, $\varepsilon_1$, are in a physically reasonable range, unlike in the previous constant capacitance model of Pokrovsky and Schott [65] or the triple layer model of Wolthers et al. [1] (they took very high values for the capacitance $C_1$, comprised between 10 and 100 F m$^{-2}$).

The partition coefficients between the Stern and diffuse layer $f_Q$ ($f_Q = -Q_\beta / Q_0$) were also computed by our surface complexation model. The partition coefficients are high and independent of pH, at a value of around 0.97, showing that most of the calcite EDL countercharge is located in the Stern layer. In the next section, the contributions of the mineral surface, Stern, and diffuse layer to the surface conductivity of calcite crystals will be computed in order to explain the origin of the high conduction currents that decrease the magnitude of the measured streaming potential.

### 3.3. Surface conductivity of calcite crystals

The surface conductivity of the calcite crystals was extract from the sample conductivity measurements using the electrical conductivity model based on the DEM theory (Eq. (13)). The estimated intrinsic formation factor was used ($F$ = 13.5) and the value of the cementation exponent $m$ was considered equal to 1.5. Results are presented at Fig. 8a. Surface conductivity is not sensitive to pH and increases with salinity. We can observe a very slight decrease of surface conductivity when pH increases except for pH >10 in the case of the lowest salinity. The Dukhin number Du was also estimated using Eq. (15) ($\mathrm{Du} = \sigma_s / 2\sigma_w$) to show the effect of crystals surface conductivity, which increases the sample conductivity and resulting conduction current (Eq. (1)), on streaming potential (Fig. 8b). The estimated Dukhin number is rather low (values <0.1) even for the lowest salinity. Therefore, one may expect no effect of surface conductivity on streaming potential. However, on the contrary to a colloidal dispersion of particles suspended in water, our calcite powder has a low connected porosity (0.18) and therefore a high quantity of conducting crystals in the sample, which increases the effects of surface conductivity on streaming potential.

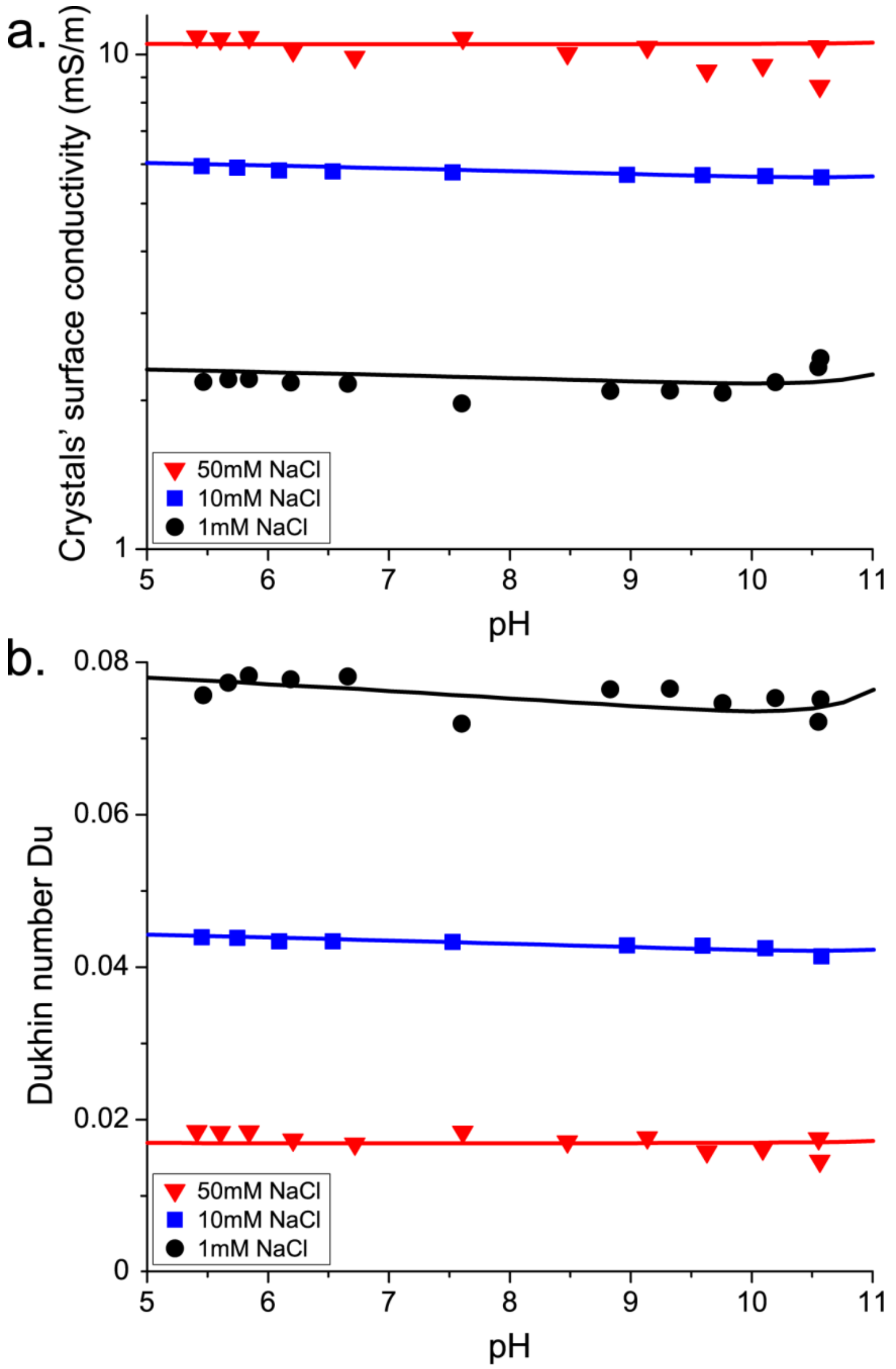

**Fig. 8.** Surface conductivity of calcite crystals (a.) and Dukhin number (b.) versus pH at different salinities (0.001, 0.01 and 0.05 M NaCl). Symbols represent the experimental surface conductivities and lines are the predictions from the BSM and surface conductivity model.

The surface conductivity effects on streaming potential and apparent zeta potential are the strongest for the lowest salinity (1 mM NaCl) where the Dukhin number is the highest ($Du \cong 0.08$; Fig. 6 and Fig. 8b). The effects of crystals surface conductivity on streaming potential decrease with increasing NaCl concentration because the bulk pore water conductivity increases more rapidly than the crystals surface conductivity with the NaCl concentration. At low salinities (for salinities typically ≤0.01 M), the ions concentration in the EDL is considerably higher than the ions concentration in the bulk water and favors electromigration currents along the particles surface. At high salinities (for salinities typically >0.01 M), the ions concentration in the bulk water become important and favors electromigration currents in the bulk water. Therefore, the Dukhin number decreases when NaCl concentration increases ($\mathrm{Du} = \sigma_s / 2\sigma_w$; Fig. 8b). To go further in the understanding of the surface conductivity of calcite particles, we decided to compute it using our surface complexation model.

The computed electrical potential at the Stern plane, surface site densities of adsorbed protons at the mineral surface and counter-ions in the Stern layer and Eqs. (14), (17)–(19), (22)–(25) were used to compute the crystals surface conductivity. The electrical potential in the diffuse layer $\varphi(x)$ was computed by numerically solving the Poisson-Boltzmann equation with Matlab using the equations developed by Leroy et al. [51]. The surface mobilities of the hydrated protons at the mineral surface ($\beta^0_{\mathrm{H_3O^+}}$) and of ions in the Stern layer ($\beta_i^{St}$) are the only fitting parameters in our surface conductivity model. Only the surface mobilities of protons and sodium ions, which are the dominating counter-ions in the Stern layer for our experimental conditions, are adjusted to decrease the number of parameters. The Simplex algorithm [64] is used to match the experimental surface conductivities with the adjusted surface mobilities. It is worth noting that the magnitude of the crystal surface conductivity depends on the value of crystal radius used in the conductivity model (Eq. (14)). A small crystal radius leads to a high surface conductivity for a given specific surface conductivity of the EDL ($\sigma_s = 2\Sigma_s / a$). This implies that the fitted surface conductivities and hence mobilities depend on the mean crystals radius chosen. In our surface conductivity model, the fitted surface mobilities decrease with the decrease of the mean crystals radius chosen. The mean crystals diameter used in our simulation was set to 2 μm (radius equal to 1 μm), which is in agreement with the micrometric sizes of calcite crystals reported in the literature [66]. Furthermore, the scaled Debye length, defined as the ratio of the particle radius to the Debye length, $a/\kappa^{-1} = \kappa a$, is largely superior to 1 ($\kappa^{-1} \cong$ 9.8 nm at the lowest salinity of $10^{-3}$ M NaCl, Eq. (24)), $\kappa a >> 1$, hence the diffuse layer can be considered thin compared to the radius of the calcite crystal.

**Table 2.** Ionic mobilities in bulk water and in the Stern layer (in $10^{-8}$ m$^2$ s$^{-1}$ V$^{-1}$; temperature $T$ = 298 K) from the Phreeqc database phreeqc.dat [56] and fitted in our study, respectively.

| Ion | $Na^+$ | $H^+$ | $Cl^-$ | $OH^-$ | $HCO_3^-$ | $Ca^{2+}$ | $CO_3^{2-}$ |
|---|---|---|---|---|---|---|---|
| $\beta_i^w$ | 5.18 | 36.25 | 7.90 | 20.52 | 4.60 | 6.18 | 7.44 |
| $\beta_i^{St}$ | 2 | 0.03 | - | - | - | - | - |

The crystals surface conductivity and Dukhin number predicted by our surface conductivity and surface complexation models are in very good agreement with the experimental data (Fig. 8a and 8b). The values of the ionic mobilities in bulk water and in the Stern layer used in our study are presented in Table 2. We found $\beta^0_{H_3O^+} = 3\times10^{-10}$ m$^2$ s$^{-1}$ V$^{-1}$ and $\beta^{St}_{Na^+} = 2\times10^{-8}$ m$^2$ s$^{-1}$ V$^{-1}$. Under the influence of the streaming potential, hydrated protons are found to migrate very slowly along the particles surface due to the dense hydrogen bonding network associated with the hydration layers. The estimated surface mobility of sodium ions in the Stern layer is quite high and confirms that sodium counter-ions are preferentially adsorbed as outer-sphere complexes at the calcite-water interface. Nevertheless, these two mechanisms must be explored further through molecular dynamics simulations in order to complete our understanding of the mechanisms responsible for the streaming potential response of calcite.

The specific surface conductivities of the mineral surface, Stern, and diffuse layer were computed to show their contributions to the total specific surface conductivity (Fig. 9). The specific surface conductivity of the mobile protons decreases when pH increases due to the deprotonation of calcium surface sites and is independent of the salinity. The Stern layer specific surface conductivity increases with the NaCl concentration because of the increasing amount of sodium ions adsorbed in the Stern layer. At a given salinity, the Stern layer specific surface conductivity increases with pH because of the deprotonation of calcium surface sites, which is in turn responsible for the decreasing negative surface charge (Fig. 9). Therefore, the surface conductivity of calcite crystals is not sensitive to pH because of the exchange of protons by sodium ions at the calcite-water interface. The Stern layer dominates the calcite surface conductivity for basic pH (pH >8) and the highest salinity (0.05 M NaCl). The diffuse layer has only a small contribution to the surface conductivity of calcite. The diffuse layer contribution decreases when salinity increases and increases with pH because of the decreasing negative surface charge and zeta potential of calcite when pH increases due to the deprotonation of surface sites (Fig. 7).

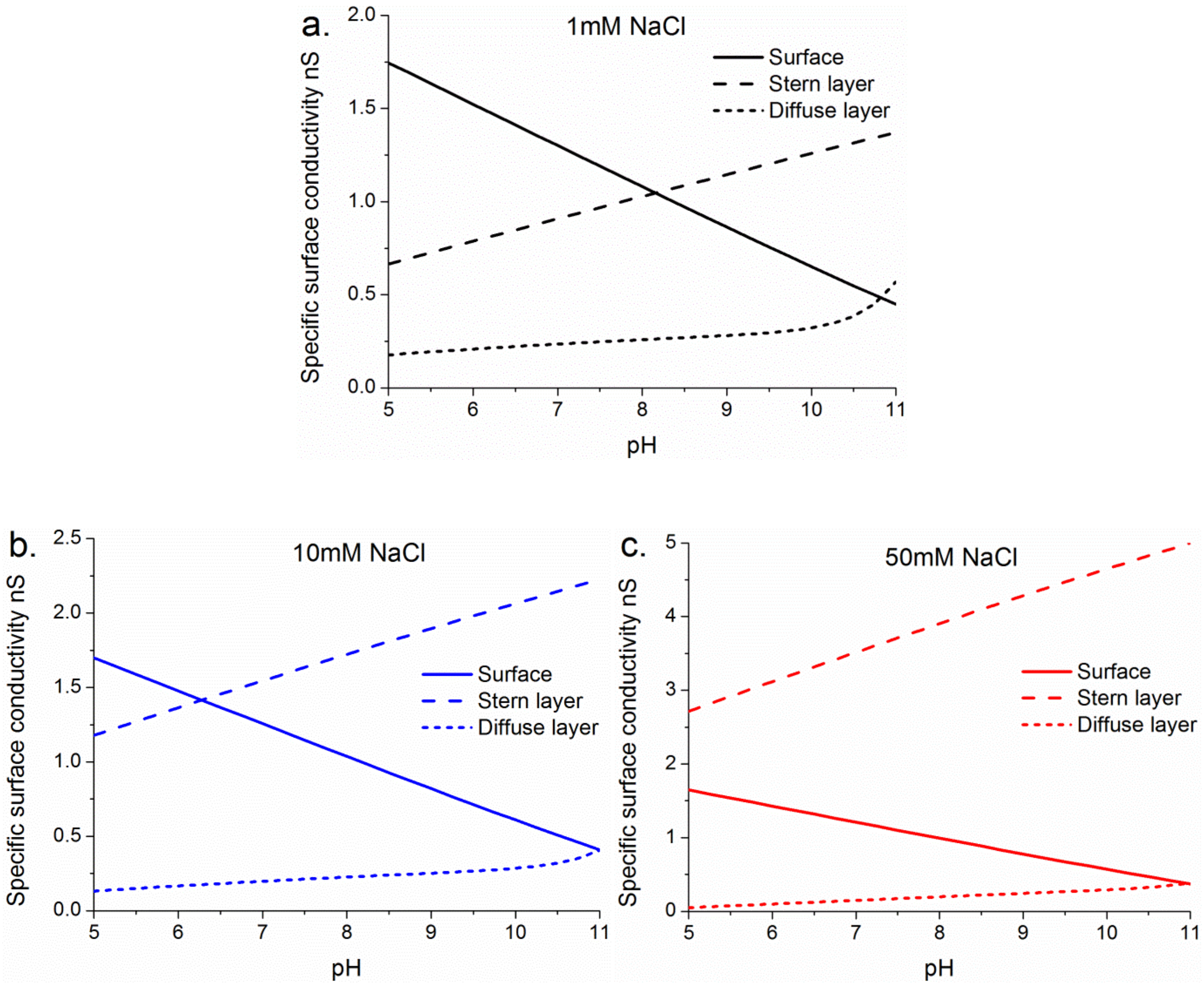

**Fig. 9.** Predicted specific surface conductivities of a calcite powder as a function of pH at three different salinities (0.001, 0.01 and 0.05 M NaCl). The specific surface conductivities are associated with the electromigration of hydronium ions at the mineral surface, sodium ions in the Stern layer, and remaining counter-ions and co-ions in the diffuse layer.

## 4. Conclusions

A new approach has been developed to characterize the electrochemical properties of calcite according to streaming potential and electrical conductivity measurements. On the contrary to the Helmholtz-Smoluchowski equation, our streaming potential model considers the influence of the particles surface conductivity and the flow regime on the streaming potential. Data measured on a calcite powder in contact with a monovalent electrolyte (NaCl) were interpreted in terms of zeta potentials and surface conductivities. The zeta potential was also computed by our basic Stern model under the assumption that the shear plane is located at the beginning of the diffuse layer. The contributions of the mineral surface, Stern, and diffuse layer to the specific surface conductivity of calcite crystals were computed as well. The following conclusions have been reached:

1. The zeta potential of calcite was easily corrected from surface conductivity and Reynolds number effects using streaming potential and electrical conductivity measurements. The corrected zeta potentials were found to be significantly larger in magnitude than the apparent zeta potentials inferred from streaming potential measurements and the Helmholtz-Smoluchowski equation, in particular when the electrolyte surrounding calcite crystals is diluted (salinity ≤0.01 M NaCl). For instance, the corrected zeta potential is approximately two and half times the value of the apparent zeta potential at a salinity of 0.001 M NaCl.
2. The surface conductivity of calcite crystals may be responsible for the discrepancy between apparent and intrinsic (corrected) zeta potentials. Surface conductivity increases the conduction current in the pores due to the streaming potential, which in turn decreases the magnitude of the streaming potential. This effect is enhanced by the surface area available for surface conduction.
3. The Helmholtz-Smoluchowski equation is not adapted to estimate the zeta potentials of calcite powders from streaming potential measurements because it neglects the effects of surface conductivity on streaming potential. The assumption of a shear plane at a certain distance from the beginning of the diffuse layer of calcite (stagnant diffuse layer assumption) may be due to the use of apparent zeta potentials to constrain the parameters of the surface complexation model and the predicted zeta potentials.

4. Our basic Stern model was able to reproduce the corrected zeta potentials without considering a stagnant diffuse layer at the calcite-water interface. Our surface complexation model predicts that most of the countercharge of the calcite-water interface (≥97%) is located in the Stern layer and that the observed independence of calcite surface conductivity on pH may result from the exchange of protons with sodium ions at the calcite-water interface.

The combination of streaming potential and electrical conductivity measurements and their interpretation using our models therefore provide a bridge to explore the fundamental processes occurring at the mineral-water interface and influencing the reactive transport properties of the porous medium. In the future, our approach can be used to describe complex conductivity measurements on carbonates, and to better understand the electrical properties of carbonates. It can also be used to better understand reactive transport phenomena, e.g. precipitation, dissolution or diffusion processes, that occur at the surfaces of carbonates.

**Acknowledgments**

This work was supported by the BRGM-Carnot Institute and the H2020 CEBAMA project. We are indebted to Dr. Mohamed Azaroual and Francis Claret for their support through the BRGM-Carnot Institute. Dr. Shuai Li post-doctoral grant is supported by the BRGM-Carnot Institute. We are grateful to Karen M. Tkaczyk for proofreading and editing the English text. We thank the two anonymous referees for the detailed reviews and the Co-Editor, Prof. Arthur T. Hubbard, for the speed and quality of the review process.

## Appendix A

The details of the calculations to correct the streaming potential coupling coefficient for the Reynolds number in the case of inertial laminar flow are presented in this Appendix. Revil and co-workers [16, 20] used the volume averaging method to upscale the transport equations at the pore scale to the transport equations at the macroscopic scale (scale of the representative elementary volume REV) in the case of inertial laminar flow (0.1 < Re < 100 [16, 20]). Their transport equations predict the total current density **J** in steady-state conditions resulting from conduction and streaming currents:

$$\mathbf{J} = -\sigma \nabla \psi - \frac{k\overline{Q}_V}{\eta_w} \nabla p, \tag{A1}$$

where $k$ is the permeability of the charged porous medium (in $m^2$) and $\overline{Q}_V$ is the volumetric excess of charge in the diffuse layer (in C $m^{-3}$).

Eq. (A1) was developed by volume averaging the local Stokes and Poisson-Boltzmann equations at the scale of a REV whatever the thickness of the EDL at the pores surface [67]. The streaming potential coupling coefficient in inertial laminar and steady-state conditions ($\mathbf{J} = \vec{0}$) is calculated according to Eqs. (4) and (A1) when the forces are directed only in the direction parallel to the water flow. The following equation is obtained:

$$C = \frac{k\overline{Q}_V}{\eta_w \sigma}. \tag{A2}$$

Teng and Zhao [21] introduced the concept of effective or apparent permeability $k$ in the case of inertial laminar flow, which is described by the following equation:

$$k = \frac{k_0}{1 + \mathrm{Re}}, \tag{A3}$$

where $k_0$ is the intrinsic permeability of rocks in the case of viscous laminar flow and Re is the Reynolds number.

The Reynolds number expresses the ratio of inertial to viscous forces in the Navier-Stokes equation [68]. It is defined according to Revil and co-workers [16, 20] by:

$$\mathrm{Re} = \frac{\rho_w \, \mathrm{U} \, \Lambda}{\eta_w}, \tag{A4}$$

where $\rho_w$ is the water volumetric density (in kg $m^{-3}$), U is the magnitude of the water velocity (in m $s^{-1}$), and Λ is a characteristic length of the flow (for capillaries Λ = *R* where *R* is the radius of the capillary).

By combining Eqs. (A2) and (A3), we obtain the following equation [16, 20]:

$$C = \frac{C_0}{1 + \mathrm{Re}}, \tag{A5}$$

where $C_0$ is the streaming potential coupling coefficient in the case of viscous laminar flow.

By analogy, Eq. (A5) is also used to relate the streaming potential coupling coefficients containing the zeta potential ($\zeta$) term (thin diffuse layer assumption) in viscous and inertial laminar conditions [16, 20] (Eqs. (5) and (9)).

The Reynolds number Re can be calculated as a function of the imposed water pressure difference and sample length, by using the reasoning described below. According to Revil [69], the length scale $\Lambda$ and the permeability $k_0$ can be described as a function of the mean grain diameter of a porous medium *d*, cementation exponent *m* and electrical formation factor *F*, by using the following equations:

$$\Lambda = \frac{d}{2m(F-1)}, \tag{A6}$$

$$k_0 = \frac{d^2}{\alpha F(F-1)^2}, \tag{A7}$$

where $\alpha$ is an empirical coefficient depending on the square of the cementation exponent ($\alpha = 32m^2$[40]). The Reynolds number is calculated as a function of the imposed water pressure difference and sample length by combining Eqs. (A3), (A4), (A6), (A7) and the Darcy equation for the water velocity (neglecting the electro-osmotic contribution), which is presented as:

$$\mathrm{U} = -\frac{k}{\eta_w}\frac{\Delta p}{l}, \tag{A8}$$

where *l* is the sample length (in m). The following equation must be solved [16, 20]:

$$\mathrm{Re}^2 + \mathrm{Re} - \frac{\rho_w}{2\alpha m \eta_w^2}\frac{d^3}{F(F-1)^3}\left(\frac{\Delta p}{l}\right) = 0. \tag{A9}$$

Finally, by solving Eq. (A9), the Reynolds number Re is calculated as a function of the imposed water pressure difference and sample length:

$$\mathrm{Re} = \frac{1}{2}\left(\sqrt{1+c} - 1\right), \tag{A10}$$

$$c = \frac{2\rho_w}{\alpha m \eta_w^2} \frac{d^3}{F(F-1)^3} \left( \frac{\Delta p}{l} \right). \tag{A11}$$

## Appendix B

A geochemical code written in Matlab was used to compute the electrical potential at the "*d*-plane" (the "*d*-plane" is assumed to coincide with the "*β*-plane" in the basic Stern model), which was considered to be equal to the zeta potential ($\zeta$). The surface site densities of adsorbed protons at the mineral surface and counter-ions in the Stern layer were also computed for the specific surface conductivity calculations. The program combines aqueous complexation, surface complexation equilibria, surface charge density and mass balance conditions. The set of equations obtained is solved iteratively by the classical Newton-Raphson technique [62].

In Table 1, the formation of the species *i* from the components *j* (solution, surface and electrostatic components) can be described using the mass action law, which is written as [11, 60]:

$$\log C_i = \sum_j n_{ij} \log X_j + \log K_i \,, \qquad \text{(B1)}$$

where $C_i$ is the concentration of species *i* (in mol dm$^{-3}$), $X_j$ is the concentration of component *j*, $K_i$ is the equilibrium constant of the formation of species *i*, $n_{ij}$ is the stoichiometric coefficient of component *j* in species *i* (i.e. the coefficients in the matrix of Table 1). The constant $\log K_i$ is related to the equilibrium formation constant of the *i*th surface or solution complex and includes the effects of pH, $pCO_2$, and the activity coefficients of dissolved species. For instance, the adsorption of cations $M^{i+}$ (from bulk aqueous solution) by $>CaOH^{-0.5}$ surface sites can be described by:

$$>CaOH^{-0.5} + M^{i+} \Leftrightarrow >CaOH^{-0.5} - M^{i+},\ K_S, \qquad \text{(B2)}$$

where $K_S$ is the equilibrium constant associated with the surface complexation reaction. This parameter can be calculated by using the following equation [11]:

$$K_S = \frac{a_{>\mathrm{CaOH}^{-0.5}-\mathrm{M}^{i+}}}{a_{>\mathrm{CaOH}^{-0.5}} a_{\mathrm{M}^{i+}}} \cong \frac{\Gamma_{>\mathrm{CaOH}^{-0.5}-\mathrm{M}^{i+}}}{\Gamma_{>\mathrm{CaOH}^{-0.5}} a_{\mathrm{M}^{i+}}} = \frac{\Gamma_{>\mathrm{CaOH}^{-0.5}-\mathrm{M}^{i+}}}{\Gamma_{>\mathrm{CaOH}^{-0.5}} a^{w}_{\mathrm{M}^{i+}}} \exp\left(\frac{q_i \varphi_\beta}{k_b T}\right), \tag{B3}$$

where $a_i$ is the activity of species *i*. The superscript "*w*" refers to ionic activities in the neutral bulk water, which is not influenced by the solid phase.

The activity of ion *i* in bulk water is determined according to:

$$a_i^w = \gamma_i^w m_i^w \cong \gamma_i^w C_i^w, \tag{B4}$$

where $\gamma_i^w$ is the activity coefficient and $m_i^w$ is the molality (in mol $\mathrm{kg}^{-1}_{\mathrm{water}}$) of ion *i*. For low to medium ionic strengths, *I* ($I \leq 0.5$ mol dm$^{-3}$ [56]), the activity coefficient is calculated by the Davies equation:

$$\log \gamma_i^w = -0.5\left(\frac{q_i}{e}\right)^2 \left(\frac{\sqrt{I}}{1+\sqrt{I}} - 0.3I\right). \tag{B5}$$

Furthermore, for each component *j*, there is an associated mass balance equation given by:

$$T_j = \sum_j n_{ij} C_i, \tag{B6}$$

where $T_j$ is the total concentration of component *j*.

For surface charge density balance conditions, the electrical potentials at the "0-plane" and "*β*-plane" are calculated using the equations describing the surface charge densities *Q* (in C m$^{-2}$). The surface charge densities at both planes are respectively computed using the following equations, which are the mass balance equations (Eq. (B6)) for the electrical potential components:

$$Q_0 = -0.5(\Gamma_{>\text{CaOH}^{-0.5}} - \Gamma_{>\text{CaOH}_2^{0.5}} + \Gamma_{>\text{CO}_3^{-0.5}} - \Gamma_{>\text{CO}_3\text{H}^{0.5}} + \Gamma_{>\text{CaOHNa}^{0.5}} + \Gamma_{>\text{CaOHCa}^{1.5}} - \Gamma_{>\text{CaOH}_2\text{Cl}^{-0.5}} - \Gamma_{>\text{CaOH}_3\text{CO}_3^{-0.5}} - \Gamma_{>\text{CaOH}_2\text{CO}_3^{-1.5}} + \Gamma_{>\text{CO}_3\text{Na}^{0.5}} + \Gamma_{>\text{CO}_3\text{Ca}^{1.5}}), \quad \text{(B7)}$$

$$Q_\beta = \Gamma_{>\text{CaOHNa}^{0.5}} + 2\Gamma_{>\text{CaOHCa}^{1.5}} - \Gamma_{>\text{CaOH}_2\text{Cl}^{-0.5}} - \Gamma_{>\text{CaOH}_3\text{CO}_3^{-0.5}} - 2\Gamma_{>\text{CaOH}_2\text{CO}_3^{-1.5}} + \Gamma_{>\text{CO}_3\text{Na}^{0.5}} + 2\Gamma_{>\text{CO}_3\text{Ca}^{1.5}}. \quad \text{(B8)}$$

The electroneutrality condition for the mineral-water interface implies:

$$Q_0 + Q_\beta + Q_d = 0. \quad \text{(B9)}$$

The surface charge density of the diffuse layer $Q_d$ is calculated as a function of the surface site densities by combining Eqs. (B7), (B8) and (B9). We obtain the following equation:

$$Q_d = 0.5(\Gamma_{>\text{CaOH}^{-0.5}} - \Gamma_{>\text{CaOH}_2^{0.5}} + \Gamma_{>\text{CO}_3^{-0.5}} - \Gamma_{>\text{CO}_3\text{H}^{0.5}} - \Gamma_{>\text{CaOHNa}^{0.5}} - 3\Gamma_{>\text{CaOHCa}^{1.5}} + \Gamma_{>\text{CaOH}_2\text{Cl}^{-0.5}} + \Gamma_{>\text{CaOH}_3\text{CO}_3^{-0.5}} + 3\Gamma_{>\text{CaOH}_2\text{CO}_3^{-1.5}} - \Gamma_{>\text{CO}_3\text{Na}^{0.5}} - 3\Gamma_{>\text{CO}_3\text{Ca}^{1.5}}). \quad \text{(B10)}$$

The surface charge density of the diffuse layer can also be calculated using the equation described below, which results from the Poisson-Boltzmann equation [18]:

$$Q_d = \sqrt{2\varepsilon_w k_b T 1000 N_\text{A} \sum_{i=1}^{N} C_i^w \left[ \exp\left( -\frac{q_i \varphi_d}{k_b T} \right) - 1 \right]}. \quad \text{(B11)}$$

The electrical potentials at the "0-plane" and "*d*-plane" are related by one molecular capacitor of capacitance $C_1$ (in F m$^{-2}$). This implies:

$$\varphi_0 - \varphi_d = \frac{Q_0}{C_1}. \tag{B12}$$

The electrical potential $\varphi_d$ is finally computed as a function of the total concentrations, equilibrium constants and capacitance $C_1$ by combining Eqs. (B10), (B11), and (B12). As shown previously, the surface site densities can be described as a function of the total concentrations, equilibrium constants and electrical potentials at the two different planes (Eqs. (B1)–(B3) and (B6)). The total concentrations, equilibrium constants, and the capacitance $C_1$ are the parameters of the surface complexation model.